-----------------------------------
Cyrus Taylor
Warren E. Rupp Assistant Professor of Science and Engineering
Department of Physics
Case Western Reserve University
Cleveland, OH   44106-7079
USA

(216) 368-3710
(216) 368-4671 (FAX)
cct@po.cwru.edu

\documentstyle{article}

\font\tenrm=cmr10
\font\tenit=cmti10
\font\elevenbf=cmbx10 scaled\magstep 1
\font\elevenrm=cmr10 scaled\magstep 1
\font\elevenit=cmti10 scaled\magstep 1

\renewenvironment{thebibliography}[1]
 { \elevenrm
   \begin{list}{\arabic{enumi}.}
    {\usecounter{enumi} \setlength{\parsep}{0pt}
     \setlength{\itemsep}{3pt} \settowidth{\labelwidth}{#1.}
     \sloppy
    }}{\end{list}}

\begin{document}
\begin{center}
\vglue 0.6cm
{
 {\elevenbf        \vglue 10pt
               OBSERVING DISORIENTED CHIRAL CONDENSATES
}
\\}
\vglue 5pt

\vglue 1.0cm
{\tenrm J. D. BJORKEN \\}
\baselineskip=13pt
{\tenit Stanford Linear Accelerator Center, Stanford University}
\baselineskip=12pt
{\tenit  Stanford, California 94309\\}

\vglue 0.3cm
{\tenrm and\\}
\vglue 0.3cm
\baselineskip=13pt
{\tenrm K. L. KOWALSKI and C. C. TAYLOR \\}
{\tenit Dept. of Physics, Case Western Reserve University}
\baselineskip=12pt
{\tenit Cleveland, OH  44106-7079\\}
\vglue 0.8cm
{\tenrm ABSTRACT}
\end{center}

\vglue 0.3cm
{\rightskip=3pc
 \leftskip=3pc
 \tenrm\baselineskip=12pt
 \noindent

We speculate that, in very high energy hadronic collisions, large
fireballs may be produced with interiors which have anomalous chiral
order parameters.  Such a process would result
in radiation of pions with
distinctive momentum and isospin distributions, and may provide an
explanation of Centauro and related phenomena in cosmic-ray events.
The phenomenology of such events is reviewed, with emphasis on
the possibility of observing such phenomena at Fermilab experiment
T-864 (MiniMax), or at a Full Acceptance Detector (FAD) at the SSC.

\vglue 0.6cm
{\elevenbf\noindent 1. Introduction}
\vglue 0.4cm
\baselineskip=14pt
\elevenrm
The vacuum is a very complicated place.  As understood in quantum
field theory, it can carry long-range order parameters,
and thus it may be possible to study coherent phenomena associated with
the alteration of the properties of the vacuum over long distances [1].

The first question one might ask is:  ``suppose that one does disorient
the vacuum over some region of spacetime.  How would one know?"  We
address this question in the context of the vacuum of the strong
interactions [2-8].  This is almost degenerate
owing to the
approximate $SU(2)_L\times SU(2)_R$ chiral symmetry.  This
symmetry is
spontaneously broken in a way analogous to what is supposed to
occur in the Higgs sector.
This phenomenon is described by the chiral fields
$$ \Phi = \sigma + i\vec \tau\cdot\vec\pi \Leftrightarrow q_L
\bar q_R,
$$
where
$$ q = \left(u \atop d\right)
$$
and
$$ \langle \Phi\rangle = \langle \sigma\rangle  = f_\pi \ne 0 \ .
$$
The fields $(\sigma,\vec\pi)$ form an $O(4)$ 4-vector.

Now suppose that in some region of spacetime the vacuum
orientation differs, and is tilted into one of the pion directions
$$ \langle \sigma\rangle  = f_\pi \cos\theta\qquad
\langle{\vec\pi}\rangle  = f_\pi \widehat u\sin\theta \ .
 $$
If we neglect the interface between this region and the exterior,
it costs relatively little energy,
$$  \Delta E = {1\over2}\, \mu^2_\pi \langle {\pi^2}\rangle = {1\over2}\,
\mu^2_\pi f_\pi^2\sin^2\theta = (10\ MeV/{fm}^3)\sin^2\theta,
 $$
arising from the pion mass in the effective Hamiltonian,
to disorient the vacuum.

How might such a region be formed?  Consider high-multiplicity, high
transverse energy collisions at collider energies.  The collision debris
from such   an interaction will expand outward at essentially the speed
of light for a considerable distance before hadronization occurs.  At
intermediate times, it is thus plausible that the geometry is that of
a `hot', relatively thin shell surrounding a `cool' interior.  (While this
picture has support even in the case of ideal hydrodynamic
expansion [9], the opposite extreme of free-streaming, without
local thermodynamic equilibrium is closer to the picture we have in mind.)
Since the interior is protected from the exterior by the hot shell, there
seems little reason to expect the interior orientation to be the same
as the exterior.

These qualitative arguments can be made a little more substantive by
appealing to a linear sigma-model description of the dynamics.  The basic
point is that if the interior is `cold', and initially in a symmetric
state, long-wavelength fluctuations will grow while short wave-length
fluctuations will be suppressed due to the instability of the meta-stable
state.  Of course, eventually the finite mass of the pion will restore the
interior to the `true' vacuum, but numerical simulations suggest that
excursions to `disoriented chiral condensates' (dcc)
may be possible [6-8].  Evidently, there are still many uncertainties
and much work to do!

Eventually, after hadronization, the surface tension will
cause the bubble of chiral condensate
to collapse.  The interior vacuum will align itself
with the vacuum of the rest of the universe, radiating pions in the process.
If the bubble of condensate is large enough to be semi-classical,
then one can study the evolution of the condensate using the equations
of motion of the sigma model. This process has a distinctive signature:
it will be coherent, and event-by-event, of a given (cartesian) isospin.
In events in which the deflection of the vacuum is in the $\pi^0$ direction,
the produced pions will be neutral; in other events in which the orientation
is orthogonal to the $\pi^0$ direction, the particles will be charged.

One can make a specific prediction for the distribution of the neutral
fraction of such pions fairly easily.  A priori, it is equally probable
that the condensate will be disoriented towards any of the (cartesian)
isospin directions.  Remembering that the fields represent probability
amplitudes, it follows that
$$P(f) df = {d f\over 2 \sqrt{f}},$$
where $f$ is the fraction of the pions which are neutral [3].

Lest one worry that we are violating conservation of isospin in this
argument, we note that it has a quantum mechanical analog.  For simplicity,
let us
consider restricting the quantum mechanical amplitude describing the
system to a single spacetime mode, and projecting onto a number basis.
If the system started as an isosinglet, for instance, then this must
be reflected in each element of the expansion of the amplitude:  each
such element must be annihilated by the isospin generators.  It follows that
such a state of (sharp) isospin zero must have an even number of pions in
it, and must be of the form
$$|\psi\rangle = C_0^{(N)}( 2 a^\dagger_+ a^\dagger_- - (a_0^\dagger)^2)^N
|0\rangle,
$$
where the $a_i^\dagger$ are the pion creation operators.  From this, it is
possible to determine the probability for seeing $2 n$ neutral pions
out of $2N$ total pions in such a state [6,10]:
$$P(n,N)={(N!)^2 2^{2N} (2n)!\over (n!)^2 2^{2n} (2N+1)!}.$$
Using Stirling's approximation, it follows that
$P(n,N) \sim 1/\sqrt{n/N}$ in the limit that bath $n$ and $N$ become
large, thus recovering the $1/\sqrt{f}$ found above.

This distribution
is significantly different from the distributions normally expected in
multiparticle production, even for fairly small values of $N$.  In
conventional multiparticle production, the distribution should be
be strongly peaked about $f=1/3$. It is is obvious that
a high percentage of dcc's
 would produce  events with anomolously small
neutral fractions; by the same token, there would also be
a significant number of events with
anomalously large neutral fractions.  We illustrate this by comparing
the isosinglet dcc distribution with a binomial distribution with
the same mean for $N=6$ in figures 1(a) and 1(b).
\vfill\eject

\vspace*{1.75in}

{\tenrm Figure 1 (a).  Probability distributions of the number of
neutral pions for total multiplicity N=6 for the binomial distribution.}
\vskip 2.5in

{\tenrm Figure 1 (b).  Probability distributions of the number of
neutral pions for total multiplicity N=6 for the isosinglet
dcc distribution.}

\vskip .25in

Before turning to the question of production mechanisms, it is
  appropriate to pause for inspiration from cosmic ray experiments.

\vglue 0.6cm
{\elevenbf\noindent 2. Are Centauro's Signals of Disoriented Chiral
Condensates?}
\vglue 0.4cm

Centauro events are cosmic ray events exhibiting [11]:

$\bullet$ Large ($ \sim 100$) numbers of hadrons;

$\bullet$ Little apparent electromagnetic energy  and hence, no
$\pi^0$'s;

$\bullet$ `High' hadronic $p_t$, reported as $k_\gamma \langle p_t \rangle
= 0.35 \pm .15\;GeV$, where $k_\gamma$ is the gamma-ray inelasticity.
\vskip.25in

In addition to the Centauro events, a class of hadron-enriched events has also
been reported [12].
Cosmic ray events with $\sum E_{tot} \geq 100\; TeV$
are presented on a scatter plot of the number of hadrons $N_h$ versus the
fraction $Q_h = \sum E_h^{\gamma} /(\sum E_h^{\gamma} + \sum E_\gamma)$
of the visible energy which these hadrons constitute.  When compared
with Monte-Carlo simulations of families based on models of the strong
interactions, and assuming that cosmic ray primaries are predominantly
protons, there are far too many ($\sim 20\%$) events in regions not populated
by the Monte-Carlo.  These events show fluctuations in hadron number and/or
energy fraction.

Before continuing, it is probably necessary to briefly review the
status of candidate Centauro events.  The 5 `classic' Centauro events
were seen in the two-storeyed emulsion chamber experiment of the
Brazil-Japan collaboration, located at 5220 m at the Chacaltaya
observatory in Bolivia.  At least two additional candidate Centauro events
have also been seen in Chacaltaya chambers.  At least one additional
candidate has been seen in the Pamir emulsion chamber experiment.  On
the other hand, it is claimed that Centauros have not been observed
in emulsion chambers at Mt. Kanbala (5500 m, China-Japan Collaboration) or
at Mt. Fuji (3750 m., Mt. Fuji Collaboration), despite comparable
cumulative exposures [13].  More precisely, the China-Japan collaboration
reports an upper limit of the fraction of such events among hadron
families with energy greater than 100 TeV to be 3\% at the 95\%
confidence level.  This appears to be a limit incompatible with the
rate at which Chacatalya has observed Centauros. This comparison may
be too glib, however, because of differences in emulsion chamber
design and data analysis. Of particular importance may be the
differing techniques for
separating hadronic showers from others.

The two groups are
similarly divided on the (non)observation of the more general class of
hadron-enriched events mentioned above.  In this context,
the debate (which has been going on for over a decade) seems to reduce
to a question of whether the data signal a change in the composition of
cosmic ray primaries at these energies, or whether they signal a change
in the hadronic interactions.

If these events are real, might they be signals of the formation of
disoriented chiral condensates?

We begin with the characteristic feature of Centauro's and the hadron-
enriched events:  anomalously large amounts of energy in the hadron
component.  The suppression of $\pi^0$'s which this implies is often
taken to indicate a suppression of pions altogether.  The argument is
basically statistical:  one would ordinarily expect the neutral fraction
to be given by essentially a binomial distribution, resulting in events
sharply peaked about $1/3$.  As we have seen above, however, the distribution
for an isospin-zero coherent state of pions is $\sim 1/\sqrt{f}$, with
$f\sim 0$ being the most probable fraction. (Note, however,
that we still have $\langle f \rangle
=1/3$.)  Thus, one is tempted to interpret the classic Centauro events as
signals of a disoriented chiral condensate.

\vspace*{7.0in}

{\tenrm Figure 2.  JACEE event [14] showing the leading particles $\eta>5$.
At lower rapidities the photon detection efficiency becomes small.}
\vfill\eject

But what about the other end of the dcc distribution, where we expect
events with an anomalously large neutral fraction?  The JACEE
collaboration has observed interesting candidate events [14], one of which we
display in Figure 2.
It is initiated by a single charged primary, and the collision occurs
within the detector.
Almost all the leading particles are photons.
The $\gamma$'s appear to cluster into two groups.
The leading cluster, indicated by the circle, consists of about
32 $\gamma$'s with $\langle{p_t}\rangle \approx$ 200 MeV and only one
accompanying charged particle.
A possibly distinct cluster has three times as many $\gamma$'s as
charged hadrons (about 54 $\gamma$'s versus about 17 charged).
This event is one out of a sample of 70 or so.
The $\gamma$/charged ratio for the generic sample is unity;
normal events are seen!
However the events are found in the emulsion by scanning for
the leading photon
showers.   So there is a ``trigger bias'' in favor of a large
neutral fraction.

There have been three systematic searches for Centauros at the CERN
collider at $\sqrt{s} = 540\; GeV$ [15],
$\sqrt{s} = 546\;  GeV$ [16], and $\sqrt{s} = 900\; GeV$ [17] and
all have yielded negative results.
The UA5 Centauro searches [16, 17] definitively ruled out
Centauros up to energies of $\sqrt{s} = 900\; GeV$; the experiment was
rather inefficient in
detecting photons, particularly in the forward direction, and
provides no limits on anti-Centauros.  The simplest reconciliation of these
experiments with the cosmic ray events is that the experimental energies
are below the threshold for the Centauro mechanism, which can be argued,
on the basis of the analysis of the cosmic ray data, to be near or
above Tevatron
energies [6, 17, 18].   Nothing like Centauros
have been seen so far at Fermilab, although a systematic search among
minimum bias events has yet to be carried out.  In this context, Goulianos
[18] has pointed out that at the Tevatron the $\eta$-distribution of a
diffractively produced Centauro is such that events of this type
would  not have been observed in the 1988-89 run, in which data were
collected requiring a coincidence between two scintillation counter arrays
covering the region $3.24 < |\eta| < 5.90 $ on both sides of the interaction
region. In any case, it appears to be very difficult for either CDF or
D0 to efficiently count both low $p_t$ charged particles and low $p_t$
photons in the forward direction as dcc interpretation of the cosmic
ray data suggests may really be required.

The upshot of all of this is that
we regard the interpretation of the cosmic ray
events in terms of dcc's as rather
speculative.  Nevertheless, we believe that this interpretation
opens an avenue towards an experimental search for this phenomena.

\vglue 0.6cm
{\elevenbf\noindent 3.  How to find Disoriented Chiral Condensates}
\vglue 0.4cm

These cosmic ray data, if they are signals of dcc's,
indicate that the signal is rather clean in the forward direction.
We now turn to the question of where one should look in an accelerator
environment, using these events as a guide.

Centauro I was close enough to the Brazil-Japan emulsion chamber that the
position of the interaction vertex could be determined by the angular
divergence of the showers in the detector.  What was measured is
$${\langle E_h^{(\gamma)} R_h\rangle\over H} = k_\gamma \langle p_t \rangle
=0.35 \pm 0.15 \; GeV,$$
where $E_h^{\gamma}$ is the portion of a hadron's energy which
is converted to (visible) electromagnetic energy, $R_h$ is the distance
of the hadron from the center of the event, and $H$ is the
height of production of the hadron.  In order to determine $p_t$,
however, one needs
to know the value of the gamma-ray inelasticity, $k_\gamma$.
The value of $k_\gamma$ is usually
quoted  as
$k_\gamma \sim 0.2-0.4$,
with the lower end being preferred for nucleons, while the
higher end is preferred for pions.  Direct measurement of $k_\gamma$ in
emulsion chambers is impossible because of the high energy threshold
($\sim 1\; TeV$).  As a result, estimates are based on extrapolating
accelerator data or on Monte-Carlo simulations.  These seem to indicate
that one should use $k_\gamma \sim 0.4$ or larger [19].
We would thus
estimate that $\langle p_t \rangle \sim 0.875 \pm 0.375\; GeV$,
with large systematic
uncertainty.

 Note that most analyses of the Centauro events
have followed the Japan-Brazil collaboration
and have used $k_\gamma = 0.2$, based on the assumption that the hadrons
are nucleons.

So where should one look for such events at the Tevatron?  To really
answer this question, one needs a better understanding of the overall
event structure than is possible either from our theoretical speculations
or from the Centauro data.  In particular, we know essentially nothing
about the central-region production
of Centauros since the Chacaltaya detector
is only sensitive to hadrons with typical hadronic transverse momenta
for
pseudorapidities of $\eta \sim 9$ or greater.
However, we can use the data to put upper
limits on the central rapidity a Centauro fireball would have at the Tevatron.

For this purpose, we assume that Centauros are diffractive fireballs,
recoiling against a proton or antiproton. This interpretation
(with, however, the identification of the hadrons as nucleons) has been
argued by K. Goulianos [18]. Under such assumptions, the center of
the Centauro fireball would be located at
$$\theta \sim {\langle p_t \rangle\over (E/\langle N\rangle )}$$
where $E=900 \; GeV$ is the Tevatron beam energy and $\langle N\rangle=
75$ is the estimated mean multiplicity of the Centauro fireballs.  Using
the above estimate for $p_t$ based on the assumption that the fireball is
composed of pions, we find that the central rapidity of a
Centauro fireball at the Tevatron is
$$\eta_c \approx 3.3 \pm 0.5.$$

This number is essentially consistent with the JACEE event which we have
suggested as an example of an anti-Centauro fluctuation.  In this event,
the photons have $\langle p_t\rangle \sim 0.2 \; GeV$.  The rapidity
densities are comparable to those of the Centauro events, and thus
we would estimate from this event that
$\eta_c \sim 4.1$ were it produced at the Tevatron.

Thus we conclude that in order to definitively search for Centauro
events at the Tevatron, under the assumption that they represent signals
of dcc's, one should be sensitive to fireballs of rapidity at least as
high at $\eta\sim 4$.

Having discussed $where$ one should look, we next turn to the question
of $how$ one should look for these events.

A useful way to assess the sensitivity of the dcc Centauro
signature is to calculate the probability of observing an event with
neutral fraction $f$
smaller than some fixed value.  For the anti-Centauro signature,
one calculates the probability for observing  $f$
larger than some fixed value. In both cases, it is useful to
study this as a function of the total pion multiplicity.
Any contamination from conventional processes (which we illustrate
using the binomial distribution) will fall off exponentially with increasing
$N$, while the disoriented condensate part will give a constant
contribution.  So in principle one would search for a break in the falling
exponential.

\vskip 2.5in

{\tenrm Figure 3.  $Log_{10}$ of the probability of finding
anti-Centauro-like configurations
($ f > 0.9$) as a function of total multiplicity $N$. The top line
represent $P(f>0.9)$ for a dcc while the lower line depicts $P(f>0.9)$
for the binomial distribution.}
\vskip 0.25in

One observation which is useful in assessing how large the acceptance of
a detector should be is that both the Centauro events and the
JACEE event have multiplicity densities which are large compared compared
to those of typical hadronic events.  Just cutting on multiplicity is
likely to enrich the sample rather dramatically.

\vglue 0.6cm
{\elevenbf\noindent 4. T-864 (MiniMax): A Chiral Condensate Search at
the Tevatron}
\vglue 0.4cm

The highest hadron
collider energies are essential.  However, at the Fermilab Tevatron, it appears
to be very difficult for CDF or D0 to  count both low-$p_t$
charged particles and low-$p_t$ photons efficiently,
particularly in the far-forward
direction. This experiment was part of the agenda of a proposal for a
Maximum Acceptance Detector (``MAX", Fermilab P-864, Bjorken and Longo
co-spokesmen [20]) at the Tevatron which, however, was rejected.
 Consequently, a small test program to initiate the study of
this physics (T-864, ``MiniMax", J. Bjorken and C. Taylor
co-spokesmen [21]) has been proposed and approved for installation
at the C0 collision area of the Tevatron during the current summer shutdown.

T-864 is a simple, staged test program of very modest scope, cost,
and impact on the laboratory which responds to the suggestion of
the Fermilab Physics Advisory Committee that it ``hopes that efforts
will continue to develop possible methods for exploring the large
rapidity regime." The experiment will initially  investigate the background
environment in the forward direction at the C0 collision area with
a minimal ``maximum acceptance" detector (MiniMax).  This will be
done initially in noncollider mode in the far forward direction
using only MWPC tracking elements and a simple scintillator-based
triggering system.   Pending successful completion of
the initial test program, MiniMax will then carry out studies of charged
particles and gamma rays in the fiducial region defined by the
MWPC telescope.  Physics goals include both generic multiparticle
production studies and, of course, a disoriented chiral condensate search.
Both studies will proceed in non-collider mode initially,  with requests
for  short collider runs following the successful completion
of the initial program.  (Under normal operating conditions of
the Tevatron,
electrostatic separators prevent collisions from occurring in
the C0 hall.)

Because of the aggressive schedule (proposed in April, approved in May,
installed by October of this year), limited resources, and constraints
from the C0 physical environment, MiniMax is designed to be extremely
flexible, in the hope that we will be able to opportunistically exploit
short unscheduled shutdowns of the Tevatron to modify the apparatus.
Figure 4 is a schematic diagram of the MiniMax detector.

\vfill\eject
Put MiniMax picture on this page
\vfill\eject

The heart of the detector is a telescope of
approximately 12 multi-wire proportional
chambers with wire spacings of $2.5\;mm$
and with an active area of about $32 \;cm$ x $32\; cm$, pointed towards the
nominal collision vertex, located 5--6 meters away from the collision
point.  Approximately 2 radiation lengths
of converter will be placed
midway throught the telescope.  This converts $80\%$ of the incident
photons while leaving $93\%$ of the charged tracks noninteracting
and  should be adequate for the statistical study of the
dcc's.

The MWPC's will be mounted
on frames which allow easy change of position both along the beam pipe,
and in position and orientation relative to the beam pipe.  This
flexibility should enable us to understand
the background environment, and will also permit us to survey a larger
range of pseudorapidities than a fixed geometry would allow.

In first
approximation, each plane will have a different orientation, permitting
the efficient use of a Hough transform algorithm for reconstructing
charged tracks and locating the vertex.  Monte-Carlo simulations
indicate that this strategy is  robust in the presence of
backgrounds arising from showers originating in the beampipe near
the detector elements.

The triggering system builds on the experience of E-735, which previously
occupied the C0 collision hall.  In addition to scintillator placed
$2 \; m$ upstream and downstream of the collision point,
there will be scintillator hodoscopes before and in the middle of the
tracking telescope.  There will also be a lead-scintillator stack at
the back of the telescope.  The trigger logic is designed to be versatile;
the rough intention is to operate with as loose a trigger as possible.
The total channel count for the trigger is approximately 70.

How sensitive might we be?  Monte-Carlo studies based on Pythia indicate
that events such as the JACEE event are extremely unusual
according to conventional wisdom, even as seen
by such a limited acceptance as that of the MiniMax detector.  If such
events occur, and if the MiniMax detector performs well and can be well
understood, it should be possible to draw significant conclusions from the
data.  A sensitivity limit of one such event per 10,000 minimum bias
events would seem to be a conservative goal.

\vglue 0.6cm
{\elevenbf\noindent 4. Future Directions}
\vglue 0.4cm

It is clear that there is a good deal of additional work to be done, both
experimentally and theoretically.

On the theoretical side, while the plausibility of the formation of
chiral condensates has been explored in various idealized situations [22-26],
there is substantial room for improvement.  Even within the context of
the various idealizations, the implications of the finite pion mass
have not been adequately investigated, nor has the question of estimating
the domain size of chiral condensate been fully
answered.  While the
long-term development of this physics will necessarily be data driven,
there is much that can and should be done now.

On the experimental side, there are a number of directions which should
be pursued in the coming years.  While MiniMax will be able to do an
initial survey of the large rapidity regime at the Tevatron, it
may make sense to follow this up with a Maximum Acceptance
Detector in the future.
Similar studies should be complemented by work in the large rapidity regime
at RHIC, where the conditions of the cosmic ray events
may be more nearly matched in proton-nucleus collisions.  Finally,
we believe that it is extremely important that large-rapidity
physics issues such as those considered here be on the agenda of
the SSC and LHC.  Indeed,
most of the work described
in this paper began in support of the inititiative to build a
Full Acceptance Detector (FAD) at the SSC.

The large-rapidity regime is one in which we have few reliable theoretical
tools and insufficient data, but what we have
learned so far is extremely tantalizing.
Onward!

\vglue 0.5cm
{\elevenbf \noindent 5. Acknowledgements \hfil}
\vglue 0.4cm
This work owes a lot to many people.  We would be dead in the water
without our T-864 collaborators, A. Abashian, P. Colestock,
F. Cverna, K. DelSignore, D. Haim
E. Kangas, W. Fickinger, R. Gustafson, B. Hanna,
T. Jenkins, L. Jones, M. Knepley, M. Longo,
M. Martens, N. Morgan, S. Oh, W. Walker, and A. Weidemann.  We are
also extremely grateful to everyone with whom we have interacted  at
Fermilab, including, in particular, C. Hojvat,
J. Peoples, J. Streets, and
T. Yamanouchi.  We have also benefitted greatly from interactions with
our colleagues in the working group for a Full Acceptance Detector (FAD)
at the SSC.  We especially thank L. Frankfurt, J. Iwai,
I. Sarcevic and M. Strikman
for their contributions.  Finally, we would like to thank the organizers
of the Madison and Argonne conferences: the final draft of the MiniMax
proposal was completed during the Madison conference, while the Argonne
conference coincided with the beginning of actual work on MiniMax!

This work was supported in part by the Department of Energy,
contract DE-AC03-76F00515; the National Science Foundation, grant
NSF-PHY-9208651, by Case Western Reserve University and by the Case
Industrial Problems and Optimization Group.
\vfill\eject

\vglue 0.5cm
{\elevenbf\noindent 6. References \hfil}
\vglue 0.2cm

}
\end{document}